# Topological Mechanics of Micromorphic Metamaterials

Mohamed Shaat

*Mechanical Engineering Department, Abu Dhabi University, Al Ain, P.O.BOX 1790, United Arab Emirates.*

E-mail address: mohamed.i@adu.ac.ae ; shaatscience@yahoo.com (M. Shaat)

**Abstract**

The topological mechanics is a perfect tool that can bridge the gap between the quantum and Newtonian physics and mechanics of materials. It requires discrete models of the material with analogies with the topological characteristics of quantum matter. Despite bridging this gap using continuous models of matter would seem challenging, we demonstrate here this possibility. We demonstrate the bridging of the quantum-continuum mechanics gap by studying the topological mechanics of metamaterials based on the micromorphic theory. A general micromorphic model of metamaterials is developed and used to study their topological characteristics. The conditions of the band-gaps, band localization, and band inversion are defined. Whereas closing the band-gap of topological insulator is obviously due to band localization, we demonstrate a mechanism of closing the band-gap with no band localization. In addition, despite the fact that the band inversion of many topological insulators cannot be completed without closing the gap, we demonstrate a case of optical band inversion with no gap closing. We foresee that the exceptional topological characteristics of metamaterials explored here will help in the design of advanced topological insulators, and open new venues of the implementations of topological insulators in mechanical applications.

## 1. Introduction

The advancement in the quantum matter with exotic properties has led to the discovery of topological insulators[1,2], photonics[3–5], phononics[6], and time non-reversal physical systems[7–15]. Despite the discrepancies between the quantum and the Newtonian descriptions of materials, a connection between them can be established based on the material's topological characteristics[16]. This has led to the topological mechanics to bridge the quantum–Newtonian mechanics gap. The topological mechanics depends on building analogies with the topological characteristics of quantum matter[17,18]. These topological analogies can be achieved using advanced artificial materials and structures[13,17–20]. The metamaterial is one potential candidate that can mimic topological insulators in achieving mechanical deformation immunity and/or non-reciprocity[13,17–21]. Therefore, a great interest is being devoted to establish the topological mechanics of various metamaterials[16–22].

Topological metamaterials can be designed by tailoring the material microstructure to allow the activation of certain nontraditional microscopic phenomena. For instance, a gyroscopic metamaterial with topologically protected unidirectional edge modes was developed by combining the gyroscopic effect and a special geometrical lattice[17]. In addition, origami and Kirigami metamaterials exhibited a topological polarization owing to their periodically folded structure[20]. Furthermore, topological metamaterials with static non-reciprocity have been realized using nonlinear elastic materials accompanied by geometrical asymmetries[13]. The analogy of the metamaterials with the topological insulators have been demonstrated by modeling the metamaterial as a mass-spring mechanical system[13,22–26] or a periodic structure[18,27,28]. These models helped in bridging the quantum–Newtonian mechanics gap by providing insightful understanding of the topological mechanics of metamaterials. Nonetheless, models that treat metamaterials as continuous matter would be rather preferred. Despite the gap between the quantum mechanics and the continuum mechanics of materials would seem unbridgeable, advanced models of continuum mechanics can bride this gap. In this paper, we study the topological mechanics of metamaterials based on the micromorphic continuum mechanics, for the first time.

The trigger of the deformation immunity and/or non-reciprocity of the metamaterial is the activation of one/more of the microstructural degrees of freedom and the inhibition of others[13,17–21,29]. Therefore, a model of the metamaterial should incorporate measures of these degrees of freedom. In the context of the micromorphic theory, the material is a multiscale material with independent microstructures, and each microstructure exhibits 12 independent-degrees of freedom[30–34]. This what gives the micromorphic theory the potential to model metamaterials[30–32,35]. In this study, we demonstrate the bridging of the quantum-continuum mechanics gap by studying the topological mechanics of metamaterials based on the micromorphic theory. We developed a general micromorphic model of metamaterials. Then, we used this model to study the topological characteristics of metamaterials. We defined the conditions that would result in the formation of the band-gap, band localization, and band inversion.

## 2. Micromorphic Model of Metamaterials

Consider a metamaterial with a large number of deformable unit cells (Fig. 1). The unit cell consists of particles, each of which is a mass-point. The unit cell deforms such that its particles independently move between two different configurations of the unit cell (Fig. 1). The unit cell rigidly moves if each two particles vibrate in-phase such that $\mathbf{w}(\mathbf{x}, t) = \mathbf{v}(\mathbf{x}, t)$, where $\mathbf{w}(\mathbf{x}, t)$ and $\mathbf{v}(\mathbf{x}, t)$ are the independent displacements of the two particles. However, if the two particles vibration such that $\mathbf{w}(\mathbf{x}, t) \neq \mathbf{v}(\mathbf{x}, t)$, the unit cell exhibits a deformation field. Therefore, the metamaterial can be modeled based on the micromorphic theory by assuming displacement and micro-deformation fields to represent the unit cell's kinematical degrees of freedom. The displacement field represents the rigid motion of the unit cell while the micro-deformation field represents the rigid-rotation and the strain of the unit cell. These two fields can be defined in terms of the independent displacements, $\mathbf{w}$ and $\mathbf{v}$, as follows (see Fig. 1):

$$\mathbf{u}(\mathbf{x}, t) = \frac{1}{2}(\mathbf{v} + \mathbf{w})$$
$$\boldsymbol{\varphi}(\mathbf{x}, t) = \frac{\mathbf{v} - \mathbf{w}}{d\mathbf{x}}$$
(1)

where $d\mathbf{x} = \mathbf{x}_Q - \mathbf{x}_P$, and $t$ denotes time.

In the context of the micromorphic theory[30,31], the displacement field and the micro-deformation field ($\mathbf{u}$ and $\boldsymbol{\varphi}$) represent 12 degrees of freedom of the unit cell (the micro-deformation field is independent of the deformation field of the entire metamaterial). Employing these degrees of freedom, the mechanics of the metamaterial is modeled based on the micromorphic theory, as follows:

| | |
|---|---|
| 1: Kinematical Variables: | |
| $\mathbf{s} = \frac{1}{2}(\boldsymbol{\varphi} + \boldsymbol{\varphi}^T)$ | Microstrain Dyadic |
| $\boldsymbol{\gamma} = \nabla \mathbf{u} - \boldsymbol{\varphi}$ | Relative Deformation Dyadic |
| $\boldsymbol{\chi} = \nabla \boldsymbol{\varphi}$ | Micro-deformation Gradient Triadic |
| 2: Equations of Motion: | |
| $\nabla \cdot \boldsymbol{\tau} + \mathbf{f} = \rho \ddot{\mathbf{u}}$ | |
| $\nabla \cdot \mathbf{m} + \boldsymbol{\tau} - \mathbf{t} + \mathbf{h} = \rho_m J \ddot{\boldsymbol{\varphi}}$ | |
| 3: Constitutive Equations (for isotropic-linear elastic metamaterials): | |
| $\mathbf{t} = \mathbf{b} : \mathbf{s} + \mathbf{d} : \boldsymbol{\gamma}$ | Microstress Tensor |
| $\boldsymbol{\tau} = \mathbf{a} : \boldsymbol{\gamma} + \mathbf{d} : \mathbf{s}$ | Cauchy-type Stress Tensor |
| $\mathbf{m} = \mathbf{c} \vdots \boldsymbol{\chi}$ | Double-stress Tensor |

(2)

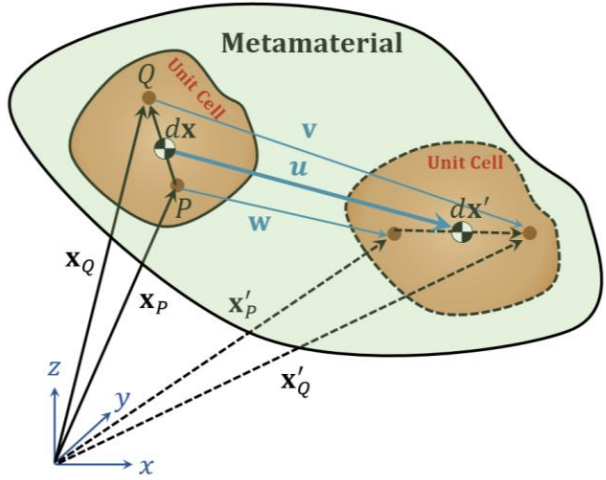

**Figure 1:** A schematic of a micromorphic model of metamaterials. The metamaterial consists of a large number of unit cells. The unit cell is a volume element that can rigidly move and deform. The displacement and deformation fields of the unit cell can be determined in terms of the displacements of two particles belong to the unit cell. Based on the given geometry, $\mathbf{v} = \mathbf{x}'_Q - \mathbf{x}_Q$, $\mathbf{w} = \mathbf{x}'_P - \mathbf{x}_P$, $d\mathbf{x} = \mathbf{x}_Q - \mathbf{x}_P$, $d\mathbf{x}' = \mathbf{x}'_Q - \mathbf{x}'_P$, $\mathbf{u} = (\mathbf{v} + \mathbf{w})/2$, and $\boldsymbol{\varphi} = (d\mathbf{x}' - d\mathbf{x})/d\mathbf{x}$.

where $\rho$ is the mass density of the metamaterial, $\rho_m$ is the mass density of its microstructure, and $J$ is a micro-inertia. $\mathbf{f}$ and $\mathbf{h}$ are body force vector and body moment tensor, respectively. $\nabla$ is the gradient operator. $\mathbf{a}$, $\mathbf{b}$, $\mathbf{c}$, and $\mathbf{d}$ are tensors of the material coefficients. For isotropic-linear elastic materials, the material coefficients are[33,36]:

$$a_{ijkl} = \lambda \delta_{ij}\delta_{kl} + \mu \delta_{ik}\delta_{jl} + \mu\eta \delta_{il}\delta_{jk}$$
$$b_{ijkl} = \lambda_m \delta_{ij}\delta_{kl} + \mu_m \delta_{ik}\delta_{jl} + \mu_m \delta_{il}\delta_{jk}$$
$$d_{ijkl} = \lambda_c \delta_{ij}\delta_{kl} + \mu_c \delta_{ik}\delta_{jl} + \mu_c \delta_{il}\delta_{jk}$$
$$\begin{aligned}c_{ijklmn} =\ & c_1(\delta_{ij}\delta_{kl}\delta_{mn} + \delta_{in}\delta_{jk}\delta_{lm}) + c_2(\delta_{ij}\delta_{km}\delta_{ln} + \delta_{ik}\delta_{jn}\delta_{lm}) + c_3\delta_{ij}\delta_{kn}\delta_{lm} \\ & + c_4\delta_{il}\delta_{jk}\delta_{mn} + c_5(\delta_{ik}\delta_{jl}\delta_{mn} + \delta_{im}\delta_{jk}\delta_{ln}) + c_6\delta_{ik}\delta_{jm}\delta_{ln} \\ & + c_7\delta_{il}\delta_{jm}\delta_{kn} + c_8(\delta_{im}\delta_{jn}\delta_{kl} + \delta_{in}\delta_{jl}\delta_{km}) + c_9\delta_{il}\delta_{jn}\delta_{km} \\ & + c_{10}\delta_{im}\delta_{jl}\delta_{kn} + c_{11}\delta_{in}\delta_{jm}\delta_{kl}\end{aligned} \quad (3)$$

where $\lambda_m$ and $\mu_m$ are the Lamé moduli of the unit cell. These moduli depend on not only the unit cell material but also the unit cell geometry[30,32]. $\lambda$, $\mu$, and $\eta$ are the coefficients of the material confined between two unit cells[30]. $\lambda_c$ and $\mu_c$ are special coefficients that define the coupling between the deformation of the unit cell and the deformation of the entire metamaterial[30]. $\delta_{ij}$ is the Kronecker delta

## 3. Topological Mechanics of Micromorphic Metamaterials

To study the topological mechanics of metamaterials, a model that deals with the material as a periodic quantum structure is needed. Based on what we propose, this periodic structure, however, should be derived from the micromorphic continuum model. For the sake of study and without loss of generality, the metamaterial is considered a one-dimensional periodic structure of deformable unit cells. Therefore, according to equation (2), the equations of motion of such a metamaterial can be derived in the form (neglecting body forces):

$$E\nabla^2 u + (E_c - E)\nabla s = \rho \ddot{u}$$
$$\xi\nabla^2 s + (E - E_c)\nabla u + (2E_c - E_m - E)s = \rho_m J \ddot{s} \qquad (4)$$

where $\nabla \blacksquare = d\blacksquare/dx$. In equation (4), $u$ is the axial displacement, and $s$ is the axial strain of the unit cell of the metamaterial. $E = \lambda + 2\mu$, $E_c = \lambda_c + 2\mu_c$, and $E_m = \lambda_m + 2\mu_m$ are equivalent elastic moduli, and $\xi$ is a material modulus with the dimension of force.

Then, equation (1) is substituted into equation (4) to give the equations of motion in the form:

$$\frac{E}{2}\nabla^2(w+v) + \frac{2(E_c - E)}{a}\nabla(v-w) = \frac{m}{Aa}(\ddot{w}+\ddot{v}) \qquad (5)$$

$$\frac{2\xi}{a}\nabla^2(v-w) + \frac{(E-E_c)}{2}\nabla(w+v) + \frac{2(2E_c - E_m - E)}{a}(v-w) = \frac{m}{A}(\ddot{v}-\ddot{w}) \qquad (6)$$

where $w(x,t)$ and $v(x,t)$ are the displacements of two particles belong to the unit cell. It should be noted that $\rho = 2m/Aa$ and $\rho_m J = ma^2/2Aa$ for a unit cell contains two particles separated by a distance $dx$, where $a = 2dx$ and $A$ are the equivalent length and cross-sectional area of the unit cell, and $m$ is the mass of one particle that belongs to the unit cell.

By the addition of equation (5) to equation (6), we obtain:

$$\left(\frac{EAa}{4} - \frac{\xi A}{a}\right)\nabla^2 w + \left(\frac{EAa}{4} + \frac{\xi A}{a}\right)\nabla^2 v - \frac{3}{4}(E-E_c)A\nabla v + \frac{5}{4}(E-E_c)A\nabla w$$
$$- \frac{(E-2E_c+E_m)A}{a}v + \frac{(E-2E_c+E_m)A}{a}w = m\ddot{v} \qquad (7)$$

In addition, by the subtraction of equation (6) from equation (5), we obtain:

$$\left(\frac{EAa}{4} + \frac{\xi A}{a}\right)\nabla^2 w + \left(\frac{EAa}{4} - \frac{\xi A}{a}\right)\nabla^2 v - \frac{5}{4}(E-E_c)A\nabla v + \frac{3}{4}(E-E_c)A\nabla w$$
$$+ \frac{(E-2E_c+E_m)A}{a}v - \frac{(E-2E_c+E_m)A}{a}w = m\ddot{w} \qquad (8)$$

The discretized model of equations (7) and (8) can be obtained by the substitution of the quantities, $\nabla^2 w = \frac{w_{n+1}+w_{n-1}-2w_n}{a^2}$, $\nabla w = \frac{w_{n+1}-w_{n-1}}{2a}$, $\nabla^2 v = \frac{v_{n+1}+v_{n-1}-2v_n}{a^2}$, and $\nabla v = \frac{v_{n+1}-v_{n-1}}{2a}$, into equations (7) and (8). This gives the equations of motion of the two particles of the *n*th unit cell, as follows:

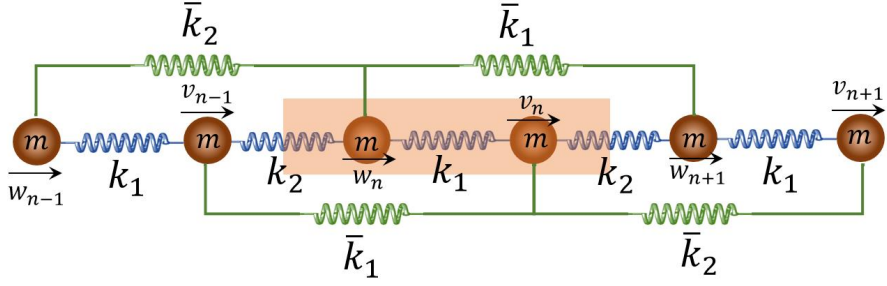

**Figure 2:** A discrete diatomic lattice model with neighbor and next-neighbor interatomic interactions that is equivalent to the micromorphic model of metamaterials.

$$m\ddot{w}_n + \frac{A}{a}\left(\frac{3}{2}E - 2E_c + E_m + 2\frac{\xi}{a^2}\right)w_n - \frac{A}{a}\left(\frac{E}{2} - 2E_c + E_m + 2\frac{\xi}{a^2}\right)v_n$$
$$- \frac{A}{a}\left(\frac{5E}{8} - \frac{3E_c}{8} + \frac{\xi}{a^2}\right)w_{n+1} - \frac{A}{a}\left(-\frac{3E}{8} + \frac{5E_c}{8} - \frac{\xi}{a^2}\right)v_{n+1} \quad (9)$$
$$- \frac{A}{a}\left(-\frac{E}{8} + \frac{3E_c}{8} + \frac{\xi}{a^2}\right)w_{n-1} - \frac{A}{a}\left(\frac{7E}{8} - \frac{5E_c}{8} - \frac{\xi}{a^2}\right)v_{n-1} = 0$$

$$m\ddot{v}_n + \frac{A}{a}\left(\frac{3}{2}E - 2E_c + E_m + 2\frac{\xi}{a^2}\right)v_n - \frac{A}{a}\left(\frac{E}{2} - 2E_c + E_m + 2\frac{\xi}{a^2}\right)w_n$$
$$- \frac{A}{a}\left(-\frac{E}{8} + \frac{3E_c}{8} + \frac{\xi}{a^2}\right)v_{n+1} - \frac{A}{a}\left(\frac{7E}{8} - \frac{5E_c}{8} - \frac{\xi}{a^2}\right)w_{n+1} \quad (10)$$
$$- \frac{A}{a}\left(\frac{5E}{8} - \frac{3E_c}{8} + \frac{\xi}{a^2}\right)v_{n-1} - \frac{A}{a}\left(-\frac{3E}{8} + \frac{5E_c}{8} - \frac{\xi}{a^2}\right)w_{n-1} = 0$$

where $w$ and $v$ are the independent displacements of the two particles, and the subscripts $n$, $n-1$, and $n+1$ refer to the unit cells $n$, $n-1$, and $n+1$, respectively.

Equations (9) and (10) indicate a diatomic lattice model with next-neighbor interactions (Fig. 2). In this model, the unit cell contains two atoms of equal masses, $m$. Atomic interactions are generated between the atom and its neighbor and next-neighbor atoms. Two different springs with stiffnesses $k_1$ and $k_2$ represent the direct-neighbor interactions. Two other springs with stiffnesses $\bar{k}_1$ and $\bar{k}_2$ represent the next-neighbor interactions. The equations of motion of the diatomic lattice model (Fig. 2) can be expressed as follows:

$$m\ddot{w}_n + (k_1 + k_2 + \bar{k}_1 + \bar{k}_2)w_n - k_1 v_n - k_2 v_{n-1} - \bar{k}_1 w_{n+1} - \bar{k}_2 w_{n-1} = 0 \quad (11)$$
$$m\ddot{v}_n + (k_1 + k_2 + \bar{k}_1 + \bar{k}_2)v_n - k_1 w_n - k_2 w_{n+1} - \bar{k}_1 v_{n-1} - \bar{k}_2 v_{n+1} = 0 \quad (12)$$

By comparing equations (11) and (12) to equations (9) and (10), the spring stiffnesses $k_1$, $k_2$, $\bar{k}_1$, and $\bar{k}_2$ are obtained in terms of the equivalent elastic moduli $E$, $E_c$, and $E_m$, as follows:

$$k_1 = \frac{A}{a}\left(\frac{E}{2} - 2E_c + E_m + 2\frac{\xi}{a^2}\right)$$

$$k_2 = \frac{A}{a}\left(\frac{7E}{8} - \frac{5E_c}{8} - \frac{\xi}{a^2}\right)$$

$$\bar{k}_1 = \frac{A}{a}\left(\frac{5E}{8} - \frac{3E_c}{8} + \frac{\xi}{a^2}\right)$$

$$\bar{k}_2 = \frac{A}{a}\left(-\frac{E}{8} + \frac{3E_c}{8} + \frac{\xi}{a^2}\right)$$

(13)

where $\xi = \frac{a^2}{8}(5E_c - 3E)$. The condition of positive spring stiffnesses requires that $E/2 < E_c < E$ and $E_m > (3E_c + E)/4$. Equations (11)-(13) represent an equivalent lattice model of the original micromorphic model of the metamaterial (Eq. (4)). It is clear that the derived lattice based on the micromorphic theory is different from the conventional diatomic lattice by employing springs with different stiffnesses along with non-neighbor interactions. In addition, the stiffnesses of the diatomic lattice are determined in terms of the elastic moduli and the microstructural geometry of the metamaterial. Therefore, we name the model in Eqs. (11) and (12) the micromorphic-diatomic lattice.

### 3.1. Dispersion relations

To derive the dispersion relations of the micromorphic-diatomic lattice, the independent displacements are expressed according to Floquet–Bloch theory, as follows:

$$w_n = W \exp(i(nq - \omega t)) \tag{14}$$

$$v_n = V \exp(i(nq - \omega t)) \tag{15}$$

where $\omega$ is the frequency, and $q$ is the nondimensional wavenumber. Then, equations (14) and (15) are substituted into equations (11) and (12) to obtain the equations of motion in the form:

$$(\Omega_1^2 + \Omega_2^2 + \bar{\Omega}_1^2 + \bar{\Omega}_2^2 - \omega^2 - \bar{\Omega}_2^2 \exp(-iq) - \bar{\Omega}_1^2 \exp(iq))W \\ - (\Omega_1^2 + \Omega_2^2 \exp(-iq))V = 0 \tag{16}$$

$$-(\Omega_1^2 + \Omega_2^2 \exp(iq))W \\ + (\Omega_1^2 + \Omega_2^2 + \bar{\Omega}_1^2 + \bar{\Omega}_2^2 - \omega^2 - \bar{\Omega}_2^2 \exp(-iq) - \bar{\Omega}_1^2 \exp(iq))V = 0 \tag{17}$$

where $\Omega_1 = \sqrt{k_1/m}$, $\Omega_2 = \sqrt{k_2/m}$, $\bar{\Omega}_1 = \sqrt{\bar{k}_1/m}$, and $\bar{\Omega}_2 = \sqrt{\bar{k}_2/m}$. The dispersion relation is obtained by setting the determinant of the coefficient matrix of equations (16) and (17) to zero, as follows:

$$\omega^2 = \left( \Omega_1^2 + \Omega_2^2 + (\bar{\Omega}_1^2 + \bar{\Omega}_2^2)(1 - \cos(q)) \right.$$

$$\left. \pm \sqrt{\frac{1}{2}(\bar{\Omega}_1^2 - \bar{\Omega}_2^2)^2(\cos(2q) - 1) + \Omega_1^4 + \Omega_2^4 + 2\Omega_1^2\Omega_2^2 \cos(q)} \right) \tag{18}$$

The dispersion relation (18) can be rewritten in the form:

$$\omega^2 = \Omega_0^2 \left( 1 + \bar{\alpha}(1 - \cos(q)) \right.$$

$$\left. \pm \frac{\sqrt{2}}{2}\sqrt{\bar{\beta}^2(\cos(2q) - 1) + \beta^2(1 - \cos(q)) + 1 + \cos(q)} \right) \tag{19}$$

where

$$\beta = \frac{\Omega_1^2 - \Omega_2^2}{\Omega_1^2 + \Omega_2^2} = \frac{2E_m + E_c - 3E}{2E_m - 4E_c + 2E}$$

$$\bar{\beta} = \frac{\bar{\Omega}_1^2 - \bar{\Omega}_2^2}{\Omega_1^2 + \Omega_2^2} = \frac{3(E - E_c)}{4E_m - 8E_c + 4E}$$

$$\bar{\alpha} = \frac{\bar{\Omega}_1^2 + \bar{\Omega}_2^2}{\Omega_1^2 + \Omega_2^2} = \frac{5E_c - E}{4E_m - 8E_c + 4E} \tag{20}$$

$$\Omega_0 = \Omega_1^2 + \Omega_2^2 = \sqrt{\frac{(E_m - 2E_c + E)A}{am}}$$

Figure 3 shows the variations of $\beta$, $\bar{\beta}$, and $\bar{\alpha}$ as functions of the microscopic modulus, $E_m$, and the coupling modulus, $E_c$. The parameters $\bar{\alpha}$ and $\bar{\beta}$, which are measures of the effect of the next-neighbor interactions on the frequencies, decrease due to an increase in $E_m$ or a decrease in $E_c$. In contrast, the parameter $\beta$, which depends on the contrast in the stiffnesses $k_1$ and $k_2$, increases due to an increase in $E_m$. A typical range of $\beta$ is $-1 < \beta < 1$ for $E_m$ varies such that $0 < E_m < \infty$.

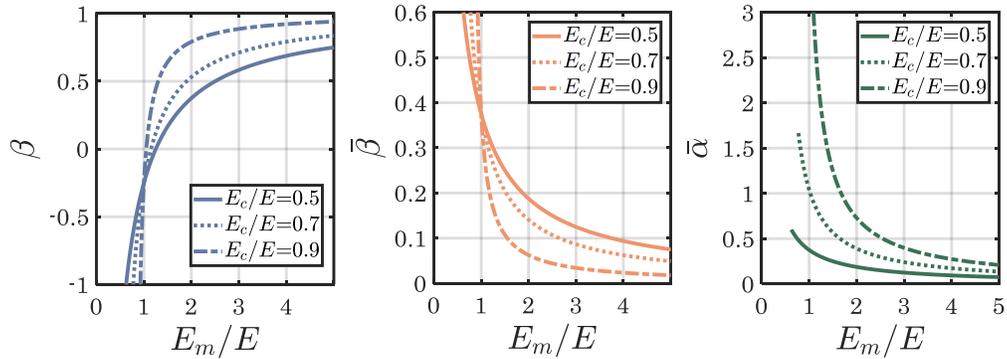

**Figure 3:** The parameters $\beta$, $\bar{\beta}$, and $\bar{\alpha}$ of the dispersion relation (Eq. (18)) versus the microstructural material modulus, $E_m$, for different values of the coupling modulus, $E_c$.

## 3.2. Band structure

Fig. 4 shows the band structures of the micromorphic-diatomic lattice for different values of $\beta = 0.5$, $0, -0.5$ and $E_c/E = 0.5, 0.7, 0.9$. Band-gaps are clearly seen in some of the obtained band structures. The band-gap width depends on not only $\beta$ but also $E_c$. When $\beta = 0$, no band-gaps are formed no matter what is the value of the modulus $E_c$. Closing the gap when $\beta = 0$ is attributed to eigenvalue loci veering of the acoustic and optical bands at $q = -\pi$ and $\pi$. When $\beta \neq 0$, band-gaps can be formed in the band structure depending on the modulus $E_c$. When $E_c/E = 0.5$, band gaps are formed as long as $\beta \neq 0$. Nonetheless, no band-gaps are observed in the band structure when $E_c/E = 0.9$. This can be attributed to the interaction between the acoustic and optical bands. As shown in Fig. 4, the optical band bends down emulating the acoustic bands as $E_c/E$ increases. The aforementioned conclusions and observations on the band-gap structure are further demonstrated as given next.

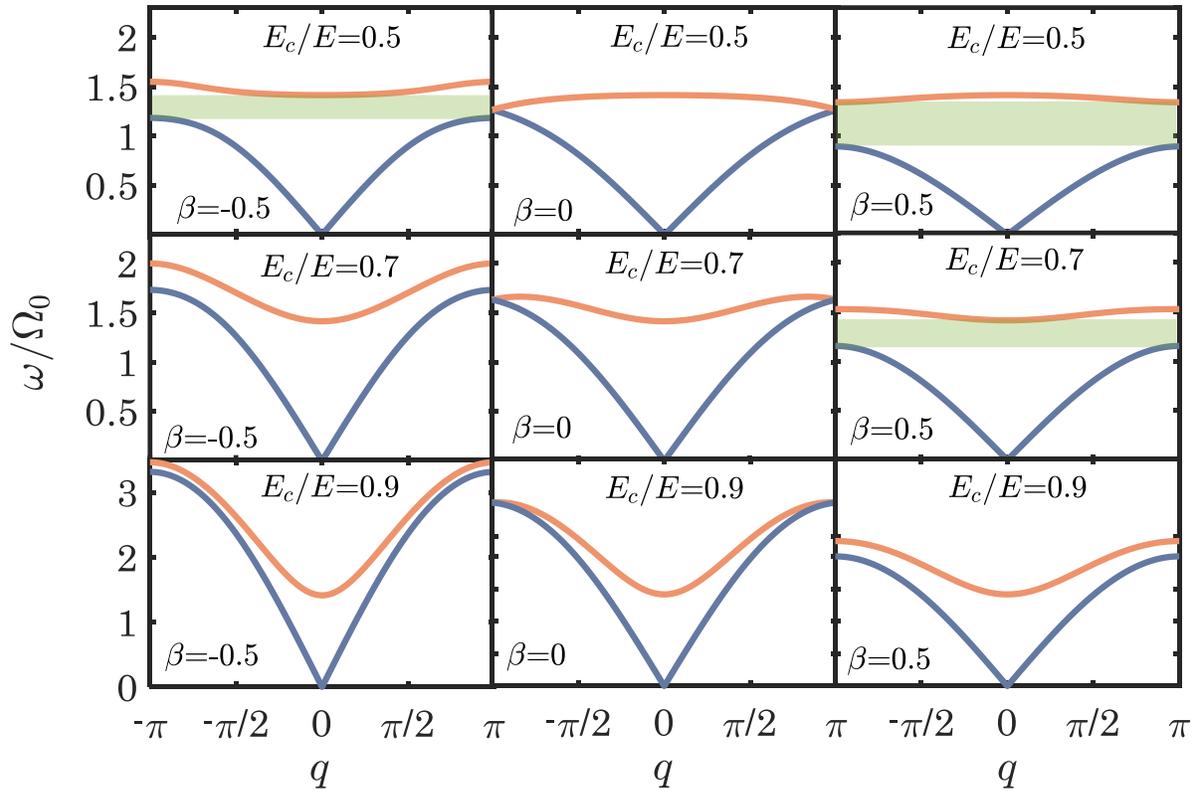

**Figure 4:** The band diagram (the nondimensional frequency ($\omega/\Omega_0$) versus the nondimensional wavenumber $q$) of the micromorphic-diatomic lattice for different values of the parameter $\beta = -0.5, 0, 0.5$ and different values of the coupling modulus $E_c/E = 0.5, 0.7, 0.9$.

## 3.3. Eigenvalue loci veering

According to equation (19), the band structure is the same regardless the value of $\beta$ is positive or negative. Thus, the dispersion relation is not enough to describe the mechanisms by which the band-gaps of micromorphic continua are formed. This – in fact – requires investigating the topological properties of these continua.

We studied the variations of the eigenvalues versus the change in the material moduli $E_m$ and $E_c$ (Fig. 5). We observed "eigenvalue loci veering" where the acoustic and optical frequencies approach each other and then veer apart as the parameter $\beta$ changes from negative to positive. It follows from Fig. 5 that the trigger of the eigenvalue loci veering is a vanishing contrast between the stiffnesses of the neighbor interactions (i.e., $\beta = 0$). Thus, the transition from "approaching" to "veering" of the acoustic and optical frequencies takes place exactly at $\beta = 0$ (or equivalently, $E_m = 3E/2 - E_c/2$).

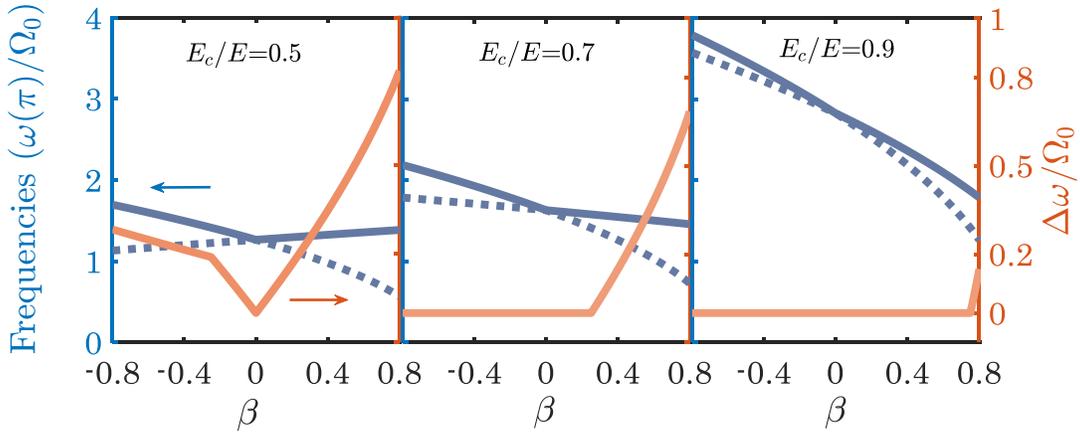

**Figure 5:** The acoustic and optical frequencies ($\omega(\pi)/\Omega_0$) and the corresponding band-gap width ($\Delta\omega/\Omega_0$) versus the parameter $\beta$. The results are represented when $E_c/E = 0.5$ (left), $E_c/E = 0.7$ (middle), and $E_c/E = 0.9$ (right).

The topological considerations indicated that the eigenvalue loci veering of micromorphic continua would require non-trivial topology or trivial topology depending on the value of the material modulus $E_c$. When $E_c/E = 0.5$, the eigenvalue loci veering is accompanied by the band-gap closing (Fig. 5). This indicates that the eigenvalue loci veering requires non-trivial topology. This result is consistent with previous observations on the eigenvalues of the classical diatomic lattices[22]. When $E_c/E \geq 0.6$, another mechanism of closing the band-gap comes into play. This mechanism mainly depends on the stiffnesses of the non-neighbor interactions (i.e., $\bar{\beta}$ and $\bar{\alpha}$). As $E_c$ increases, the stiffnesses $\bar{k}_1$ and $\bar{k}_2$ increase.

Because of the non-neighbor interactions, the optical bands bends to emulate the acoustic bands such that the optical bands become coincident with the acoustic bands when $E_c \to 1$ (see Fig. 4). Because of the bending of the optical bands, gaps would be closed even if $\beta \neq 0$ (Fig. 5). In other words, the gap would be closed without eigenvalue loci veering.

### 3.4. Band inversion and band localization

The eigenvalue loci veering is usually accompanied by a fast variation in the eigenvectors; therefore, we expect a mode (or band) inversion[37] or localization[38,39]. To investigate this, we studied the variation of the acoustic and optic eigenvectors with the variation of the material moduli, $E_m$ and $E_c$. According to Eqs. (16)-(18), the acoustic and optic eigenvectors, $\boldsymbol{\Psi}$, are determined in the form:

$$\boldsymbol{\Psi} = \{1 \quad \Psi(q)\}^{\mathrm{T}} \tag{21}$$

where

$$\Psi(q) = \frac{C(q)}{Z \mp \sqrt{Z^2 + |C(q)|^2}} \tag{22}$$

where $C(q) = \Omega_1^2 + \Omega_2^2 \exp(-iq)$ is the off-diagonal component of the dynamical matrix of the equations of motion (Eqs. (16) and (17)), and $|C(q)|$ is its magnitude. The complex number $Z = i(\bar{\Omega}_2^2 - \bar{\Omega}_1^2)\sin(q)$ depends on the non-neighbor interactions. By neglecting the non-neighbor interactions, i.e., $Z = 0$, the acoustic and optic eigenmodes, $\Psi(q)$, become:

$$\Psi(q) = \mp \frac{C(q)}{|C(q)|} \tag{23}$$

Figure 6 shows the contour plots of the acoustic and optic eigenmodes, $\Psi(q)$, in the complex plane for different values of the parameter $\beta = -0.5, 0,$ and $0.5$. These contours show the variation of the eigenmodes as the nondimensional wavenumber $q$ changes from $-\pi$ to $\pi$.

Fig. 6(a) shows the contours of the acoustic and optic eigenmodes when neglecting the non-neighbor interactions (Eq. (23)). Different contours are obtained for the different values of the parameter $\beta$. When $\beta \geq 0$, the acoustic and optical modes are distinct where the contours of the acoustic (optic) eigenmodes are confined to the right (left) side of the complex plane. The acoustic mode is in-phase ($\mathrm{Re}(\Psi(q)) > 0$), while the optical mode is out-of-phase ($\mathrm{Re}(\Psi(q)) < 0$). When $\beta = 0$, the eigenvalue loci veering occurs when $q = \pi$ and $-\pi$, and it results in a band localization, i.e., $\mathrm{Re}(\Psi(q)) = 0$ and $\mathrm{Im}(\Psi(q)) = 0$. Both the acoustic and optic bands are localized where the vibration energy is confined at only one atom of the diatomic lattice. For such a case, there is no winding about the origin, and there is no band inversion. When $\beta > 0$, the acoustic and optic eigenmodes are highly distinct for the entire range of the wavenumber. There is no winding around the origin of the complex domain, and hence no band inversion. In addition, there is no band localization where the vibration energy distributes between the two atoms of the diatomic lattice. When $\beta < 0$, the acoustic and optic contours wind about the origin of the complex plane, and the winding number is one for both bands. This indicates that both the acoustic and optic bands are inverted as $q$ changes from 0 to $\pi$ or $-\pi$. The results represented in Eq. 6(a) for the diatomic lattice without non-neighbor interactions indicate a non-trivial topology. Thus, the trigger of the inversion of the acoustic and optic bands is a vanishing gap.

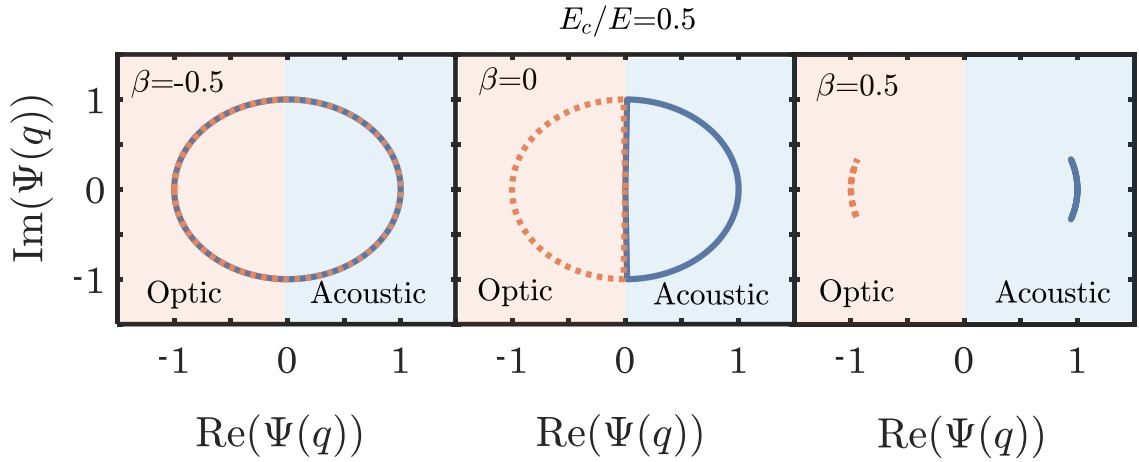

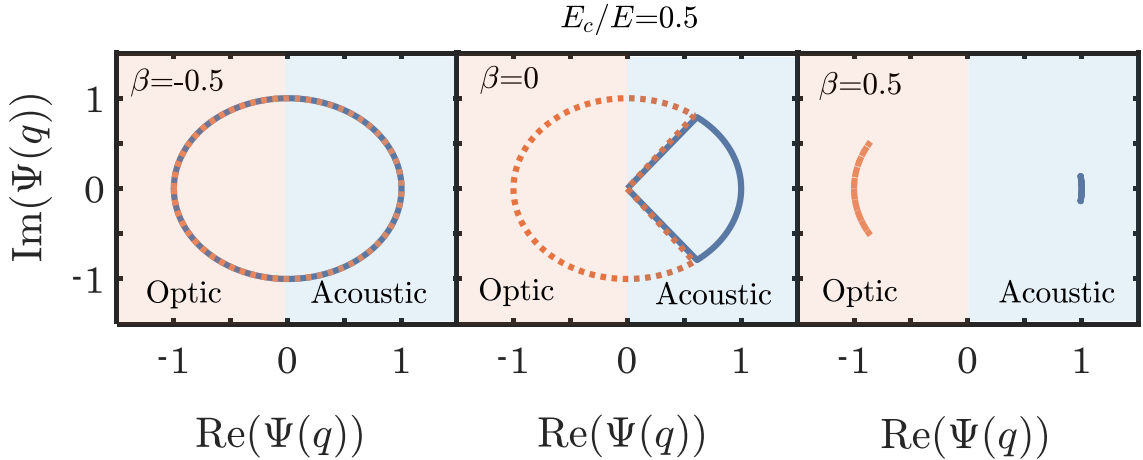

**Figure 6:** Contour plots of the acoustic and optic eigenmodes, $\Psi(q)$, in the complex plane for different values of the parameter $\beta = -0.5$ (left), $\beta = 0$ (middle), and $\beta = 0.5$ (right). **(a)** The contour plots are represented when neglecting the non-neighbor interactions, i.e., $\bar{k}_1 = 0$ and $\bar{k}_2 = 0$. **(b)** The contour plots are represented when considering the non-neighbor interactions.

The contours of the acoustic and optic eigenmodes when considering the non-neighbor interactions (Eq. (22)) are depicted in Fig. 6(b). When $\beta = 0$, both the acoustic and optic bands are localized ($\text{Re}(\Psi(q)) = 0$ and $\text{Im}(\Psi(q)) = 0$) at $q = \pi$ and $-\pi$ due to the eigenvalue loci veering. However, the acoustic and optic bands are non-localized as long as $\beta \neq 0$. The acoustic and optic bands wind once around the origin of the complex plane (i.e., the winding number is one) as long as $\beta < 0$ (no winding when $\beta \geq 0$). Whereas there is no winding when $\beta$ is positive, the optic bands would be inverted as $q$ changes from 0 to $\pi$ or $-\pi$. This is attributed to the non-neighbor interactions. Thus, the optic band inversion takes place when $\beta < 0.08$. These observations indicated that, because of the non-neighbor interactions, the optic band inversion can be achieved without closing the band-gap. On the other hand, the acoustic band inversion requires closing the gap where it is achievable as long as $\beta < 0$.

## Conclusions

Here, we demonstrated the possibility to bridge the gap between the quantum and Newtonian physics and mechanics of materials using continuous models of matter. Based on the topological mechanics of metamaterials, we established an analogy of the micromorphic mechanics with the quantum mechanics. We linked the topological characteristics of metamaterials to their mechanical properties. We demonstrated that the responsive behavior of metamaterials would be topologically non-trivial/trivial depending on its elastic moduli. In addition, we defined the conditions of the band localization and band inversion in terms of the material's elastic moduli.

We revealed exceptional topological characteristics of micromorphic metamaterials. Whereas closing the band-gap of topological insulator mainly requires band localization, we demonstrated another mechanism of closing the band-gap without band localization. This mechanism depended on a material modulus, $E_c$, that promoted the non-neighbor interactions. Because of the non-neighbor interactions, the optical bands would merge with the acoustic bands resulting in coincident bands but of different phases (when $E_c/E \to 1$). In addition, the inversions of the acoustic and optical bands of the classical diatomic lattice-based topological insulators are completely twistable and cannot be completed without closing the gap[22]. Nonetheless, the micromorphic diatomic lattice would exhibit an optical band inversion without closing the gap.

## Acknowledgements

This research is supported by Abu Dhabi University (Grants 19300474 and 19300475).

## Data Availability

The data that support the findings of this study are included in the article.

## Competing Interests

The author has no competing interests to disclose.